\documentclass[showpacs,twocolumn,prb]{revtex4-1}
\usepackage{amsfonts}
\usepackage{graphicx}
\usepackage{amsmath}
\usepackage{amssymb}

\begin{document}

\title{Particle-hole asymmetry on Hall conductivity of a topological
insulator}
\author{Zhou Li$^1$}
\email{lizhou@mcmaster.ca}
\author{J. P. Carbotte$^{1,2}$}
\email{carbotte@mcmaster.ca} \affiliation{$^1$ Department of
Physics, McMaster University, Hamilton, Ontario,
Canada,L8S 4M1 \\
$^2$ Canadian Institute for Advanced Research, Toronto, Ontario,
Canada M5G 1Z8}
\begin{abstract}
The helical Dirac states on the surface of a topological insulator are
protected by topology and display significant particle-hole asymmetry. This
asymmetry arises from a subdominant Schr\"{o}dinger type contribution to the
Hamiltonian which provides a small perturbation to a dominant Dirac
contribution. This changes the Landau levels energies in an external
magnetic field ($B$) and provides modifications to the usual relativistic
optical matrix elements. Nevertheless we find that the relativistic
quantization of the Hall plateaux remains even when the ratio of the Schr%
\"{o}dinger ($E_0$) to Dirac ($E_1$) magnetic energy scale increases
either through an increase in $B$, a decrease in the Schr\"{o}dinger
mass or of the Dirac fermi velocity. First corrections to the
optical matrix elements(OME) in the relativistic case drop out at
least to order $(E_0/E_1)^3$. In the opposite limit $E_1$ small, the
quantization remains classical but there is a split into two series.
The first corrections to the OME in this case, cancel out at least
to order $(E_1/E_0)^4$.
\end{abstract}

\pacs{73.43.Cd,71.70.Di,73.25.+i}
\date{\today }
\maketitle

\section{Introduction}

In graphene, the Dirac cones associated with conduction and valence bands
are normally taken to have perfect particle-hole symmetry about the Dirac
point. On the other hand the topologically protected helical Dirac fermions
on the metallic surface \cite{Hasan,Qi} of a topological insulator (TI) also
show Dirac dispersion \cite{Chen1,Xu,Chen2,Hsieh} curves with spin momentum
locking, but they generally display significant particle-hole asymmetry \cite%
{Chen1,Xu,Chen2,Hsieh,Hancock,Li1,Li2,Li3}. This leads to electronic
dispersion curves characterized by an hourglass or goblet shape with
valence band below the Dirac point fanning out more rapidly than the
corresponding conduction band. Its wider base eventually merges with
the bulk valence band at which point it no longer can be traced as a
separated entity. This behavior is modeled by including, in addition
to a dominant Dirac linear in momentum piece, a subdominant
Schr\"{o}dinger contribution quadratic in momentum with mass m. This
term leads directly to the restructuring of the perfect Dirac
dispersions to goblet dispersions instead.

Even a small Schr\"{o}dinger term in the Hamiltonian can have
important consequences for the properties of
topological insulators. As an example, Wright and McKenzie \cite%
{Wright1,Fuchs,Wright2} have very recently found for gapped systems
that a finite $m$ term can lead to important changes in the phase of
the quantum oscillations
associated with Shubrikov-de-Hass or de-Hass-van Alphen effects.\cite%
{Ando,Mikitik,Champel,Kopel,Sharapov,Gusynin,Wang} The phase offset ($\gamma
$) of these oscillations is determined not only by the Berry phase of the
cyclotron orbits involved, but by a further amount which exists only when
the Schr\"{o}dinger term is present i.e. $m\neq \infty $. \cite%
{Wright1,Fuchs,Wright2} As a second example, the magneto optical absorption
lines of a TI \cite{Li2} split into two peaks in contrast to the single peak
of graphene because of the particle-hole asymmetry. The interband
transitions allowed by the optical selection rules from Landau level (LL) $%
-n $ in the valence band to $n+1$ in the conduction band and from $-(n+1)$
to $n $ are no longer degenerate in energy as they would be in graphene.
\cite{Carbotte,Pound,Tabert}

In graphene the electron dynamics is determined by the relativistic Dirac
equation and the integer quantum Hall effect is unconventional. \cite%
{Haldane,Gusynin1,Gusynin2,Novo,Zhang,Liu}The DC Hall conductivity has
plateaux in units of $2e^{2}/h$ at $(2n+1)$ with $n=0,1,2,3...$ which is to
be contrasted with the conventional case of Schr\"{o}dinger dynamics where
quantization is $2n$ rather than $(2n+1)$. An important question which we
wish to address in this paper is how does a small subdominant Schr\"{o}%
dinger piece in the Hamiltonian change the Dirac sequence of the dominant
Dirac piece. Even without such a complication we know that impurity
scattering and/or temperature affects the integrity of the Hall plateaux
eventually smearing them out towards the classical unquantized result.\cite%
{Gusynin1,Gusynin2}

The paper is structures as follows. The formalism
associated with the landau levels (LL) created by an external magnetic field (%
$B$) oriented perpendicular to the surface of the topological
insulator is presented in section II. Eigen energies and wave
functions are written in terms of the Schr\"{o}dinger $E_{0}=\hbar
e|B|/m$ and Dirac $E_{1}=\hbar v_{F}\sqrt{e|B|/\hbar }$ magnetic
energies with $v_{F}$ the Dirac velocity. The general formula for
the Hall conductivity is specified, and its DC limit is taken. In
section III the resulting formula is shown to reduce to the
known Schr\"{o}dinger and Dirac quantization $0$, $1$, $2$, $3...$ and $1/2$, $%
3/2 $, $5/2...$ respectively when there is only a quadratic or
linear in momentum term in the Hamiltonian. For graphene we have a
factor of $4$ from spin and valley degeneracy not included here. In
the more general case when both terms are present the resulting
expressions are complicated. In section IV, expanding OME in powers
of $E_{0}/E_{1}$, we show that the resulting simplified equation
that determines the Hall plateaux has the same form as for pure
Dirac but with the new Landau level (LL) energies appropriate to the
TI. All corrections from OME have dropped put at least to order
cubic in the ratio $E_{0}/E_{1}$. We also consider the opposite
limit appropriate to present day spintronic semiconductors for which
the Schr\"{o}dinger term is dominant but a small Dirac contribution
is also present. Expanding the OME in powers of $E_{1}/E_{0}$ we
find that they cancel out at least to order $(E_{1}/E_{0})^4 $ and
the equations reduce to the pure classical equations but with the
new LL energies, and two such series are involved. In the same
section we present numerical results. Details of the derivations are
to be found in an appendix. A summary and conclusions make up
section V.

\section{Formalism}

\begin{figure}[tp]
\begin{center}
\includegraphics[height=4.0in,width=3.0in]{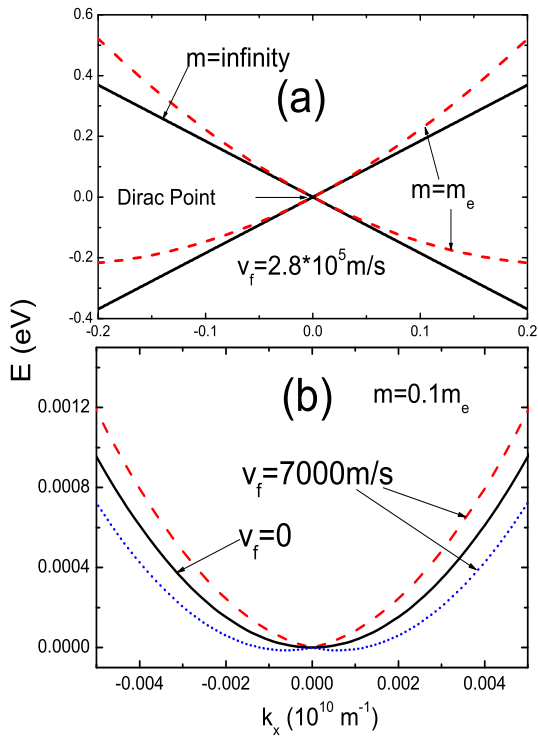}
\end{center}
\caption{(Color online) a) Solid curves (black) give the perfect Dirac cones
when the Schr\"{o}dinger piece in Eq.~(\protect\ref{H0}) is ignored. These
are shown for comparison with the dashed curve (red) which includes both Schr%
\"{o}dinger and Dirac contribution (topological insulators). b) Solid curve
(black) gives the perfect Schr\"{o}dinger dispersion when the Dirac piece in
Eq.~(\protect\ref{H0}) is ignored. This is shown for comparison with the
dashed red and dotted blue curves which include both Schr\"{o}dinger and
Dirac contribution (spintronic semiconductors).}
\label{fig1}
\end{figure}

\begin{figure}[tp]
\begin{center}
\includegraphics[height=3.0in,width=3.2in]{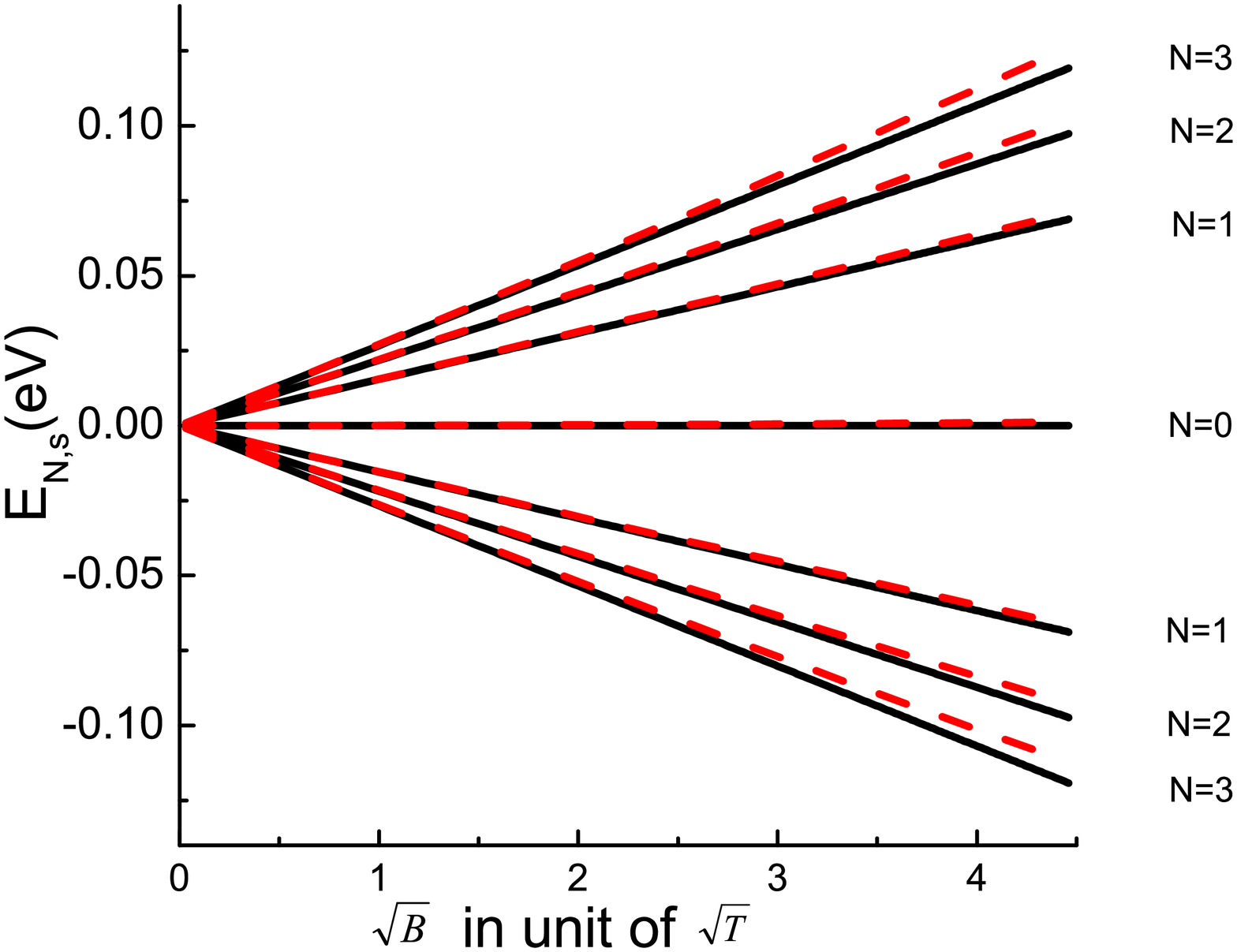}
\end{center}
\caption{(Color online) Landau level energies $E_{N,s}$ labeled by the index
$N$ as a function of the square root of the magnetic field $B$ measured in
Tesla for the same band structure parameters as in Fig.~\protect\ref{fig1}%
(a). Solid (black) straight lines are for comparison and represent the pure
Dirac case for which $m=\infty $ in Eq.~(\protect\ref{H0}). The dashed (red)
lines include a small Schr\"{o}dinger contribution. This can have a strong
effect, particularly on the negative energy state with large value of $N$
(LL index). }
\label{fig2}
\end{figure}
A minimal Hamiltonian for describing the helical Dirac fermions that exist
at the surface of a three dimensional topological insulator has the form
\begin{equation}
H_{0}=\frac{\hbar ^{2}k^{2}}{2m}+\hbar v_{F}(k_{x}\sigma _{y}-k_{y}\sigma
_{x})  \label{H0}
\end{equation}
where $\sigma _{x}$,$\sigma _{y}$ are the spin Pauli matrices and $\mathbf{k}
$ is momentum. The first term is the usual quadratic in momentum Schr\"{o}%
dinger contribution with effective mass $m$ and the second describes Dirac
fermions with velocity $v_{F}$. The dispersion curves associated with Eq.~(%
\ref{H0}) are
\begin{equation}
\varepsilon _{k,\pm}=\frac{\hbar ^{2}k^{2}}{2m}\pm\hbar v_{F}k  \label{E0}
\end{equation}
These are displayed in Fig.~\ref{fig1}a for a set of parameters typical for
topological insulators namely $v_{F}=2.8\ast 10^{5}m/s$ and $m$ equal to the
bare electron mass ($m_{e}$). For reference, in the specific case of $%
Bi_{2}Te_{3}$, $v_{F}=4.3\ast 10^{5}m/s$ and $m=0.09m_{e}$. The solid
(black) curves in Fig.~\ref{fig1}a are for pure Dirac and are included for
comparison with dashed (red) curves which give the electron dispersion $%
\varepsilon _{k,\pm}$ of Eq.~(\ref{E0}) when $m=m_{e}$. This piece adds on
to the black curves in both conduction and valence band and leads to a
goblet or hourglass shape. It narrows the cone cross-section as energy is
increased in the conduction band while it widens that in the valence band
with decreasing energy below the Dirac point at $E=0$. The particle-hole
asymmetry displayed in Fig.~\ref{fig1}a can be characterized by the value of
momentum $k_{c}$ at which the valence band has a minimum value. We find $%
k_{c}=v_{F}m/\hbar \simeq .58\mathring{A}^{-1}$  and the energy at minimum is $\frac{\hbar ^{2}k_{c}^{2}%
}{2m}-\hbar v_{F}k_{c}=-\frac{1}{2}mv_{F}^{2}\simeq -220meV$. For the
parameters estimated from first principle calculations in $Bi_{2}Te_{3}$
this minimum energy would be much smaller of order $48meV$. In Fig.~\ref%
{fig1}b we show the same dispersion curves but for a very different set of
parameters which are more representative of semiconductors presently used in
spintronics. The Dirac fermi velocity is much smaller than that used in the
top frame for topological insulators and the Schr\"{o}dinger mass has also
been taken to be a factor of 10 smaller. The solid black curve applies to
the pure Schr\"{o}dinger case and is for comparison. The red dashed and blue
dotted curves include the small Dirac contribution.

Turning next to the effect on electron dynamics of a magnetic field $B$
oriented perpendicular to the surface of the TI we replace Eq.~(\ref{H0}) in
the Landau gauge by
\begin{eqnarray}
H_{0} &=&\frac{\hbar ^{2}[(-i\partial _{x})^{2}+(-i\partial _{y}+eBx/\hbar
)^{2}]}{2m}  \notag \\
&&+\alpha \lbrack (-i\partial _{x})\sigma _{y}-(-i\partial _{y}+eBx/\hbar
)\sigma _{x}]  \label{HM}
\end{eqnarray}%
where $\alpha =\hbar v_{F}$, $\mathbf{B}=B\hat{z}$ and $\hat{z}$ is
a unit vector perpendicular to the surface plane. The energies of
the Landau levels
(LL) ignoring Zeeman splitting are%
\begin{equation}
E_{N,s}=\hbar ^{2}N/(ml_{B}^{2})+s\sqrt{[\hbar
^{2}/(2ml_{B}^{2})]^{2}+2N\alpha ^{2}/l_{B}^{2}}  \label{landau}
\end{equation}%
where the magnetic coherence length $l_{B}=1/\sqrt{e|B|/\hbar }$ with $e$
the electron charge, $N\neq 0$ is the LL index and $s=\pm $ gives conduction
($+$) and valence ($-$) band respectively. For $N=0$
\begin{equation}
E_{N=0}=\hbar ^{2}/(2ml_{B}^{2})  \label{landau0}
\end{equation}%
The energy levels $E_{N,s}$ as a function of the square root of the
magnetic field $B$ in Tesla are plotted in Fig.~\ref{fig2} for the
illustrative parameters defined in Fig.~\ref{fig1}a. It is
convenient to define a Schr\"{o}dinger magnetic energy scale as
$E_{0}=\hbar e|B|/m$ (here $E_{0}\approx0.116$ meV for $B=1$ T) and
an equivalent Dirac magnetic scale as $E_{1}=\alpha \sqrt{e|B|/\hbar
}$ (here $E_{1}\approx10.4$ meV for $B=1$ T) and also introduce a
Diracness ratio $P\equiv E_{1}^{2}/E_{0}^{2}$. In this case
$P\rightarrow \infty $ corresponds to pure Dirac and $P\rightarrow
0$ to pure Schr\"{o}dinger. The significance of this parameter is
that we will seek corrections to pure Dirac in powers of $1/P\equiv
x$ with $x<<1$ and pure Schr\"{o}dinger in powers of $P$ with
$P<<1$. While we will show later that our theory does reduce to the
well known integer quantum Hall effect seen in semiconductors, when
we consider the limit $P\rightarrow 0$, for the main part we will be
interested in small deviation from a pure Dirac
case or $E_{0}/E_{1}=1/\sqrt{P}$ small. Turning specifically to Fig.~\ref%
{fig2} the solid (black) lines give the energy levels when $E_{0}$ is taken
to be zero, i.e. $m\rightarrow \infty $ in the Hamiltonian Eq.~(\ref{H0}).
These straight lines as a function of $\sqrt{B}$ are for comparison. The
dashed (red) curves represent the case when a small but finite Schr\"{o}%
dinger contribution is included in addition to the dominant Dirac
contribution. It is clear that deviations between solid and dashed
curves increase with $N$ as they do with increasing $B$. These
deviations are much more pronounced for the negative energy Landau
levels as can be expected from Fig.~\ref{fig1} which shows that for
$B=0$, the valence band has a minimum of order $1/4$ eV which occurs
at higher values of $k$ than is shown. This is
reflected directly in the LL energies. While not seen clearly in Fig.~\ref%
{fig2} because of the restricted range of $\sqrt{B}$ and of $N$ shown, the
negative LL index curves at large N will have a minimum at some finite value
of $\sqrt{B}$, then increase and eventually cross the $E=0$ line to become
positive. Note that if we expand the magnetic LL energies in lowest power of
$E_{0}$ we get
\begin{equation}
E_{N,s}=s\sqrt{2N}E_{1}+E_{0}[N+\frac{s}{8}\frac{1}{\sqrt{2N}}\frac{E_{0}}{
E_{1}}]  \label{ENS}
\end{equation}%
The leading correction is of order $E_{0}$ which scales like $B$ and
hence shows a quadratic departure of the dashed (red) curve in
Fig.~\ref{fig2} from the solid black straight lines. The coefficient
of $E_{0}$ in Eq.~(\ref{ENS}) is $N$ so that these quadratic
deviations increase with value of LL index $N$.

The DC transverse Hall conductivity $\sigma _{xy}(\omega =0)$ can be
calculated from the LL energies of Eq.~(\ref{landau}) but further
required a knowledge of the corresponding eigen functions. Following
Ref.~[\onlinecite{Li2}] we can write
\begin{equation}
|N,s\rangle =\left[
\begin{array}{c}
C_{\uparrow ,N,s}|N-1\rangle _{\uparrow } \\
C_{\downarrow ,N,s}|N\rangle _{\downarrow }%
\end{array}%
\right]\label{wave}
\end{equation}%
with $s=+/-$ and the coefficients $C_{\uparrow (\downarrow ),N,s}$ can be
written in terms of the Diracness index $P$ previously introduced. For $N>0$%
\begin{eqnarray}
C_{\uparrow ,N,s} &=&\frac{\sqrt{\sqrt{1/4+2NP}-s/2}}{J}  \notag \\
C_{\downarrow ,N,s} &=&\frac{-s\sqrt{\sqrt{1/4+2NP}+s/2}}{J}
\end{eqnarray}%
with $J=\sqrt{2\sqrt{1/4+2NP}}$. For $N=0$, $C_{\uparrow ,0}=0$ and $%
C_{\downarrow ,0}=1$ and only $s=+$ need be considered.The current operator $%
j_{\alpha }$ is related to velocity $v_{x}$ and $v_{y}$ by
\begin{equation}
v_{x}=\frac{\hbar k_{x}}{m}+\frac{\alpha }{\hbar }\sigma
_{y}\label{vx}
\end{equation}%
and%
\begin{equation}
v_{y}=\frac{\hbar (k_{y}+eBx/\hbar )}{m}-\frac{\alpha }{\hbar
}\sigma _{x}\label{vy}
\end{equation}%
The standard Kubo formula for the finite frequency ($\omega$)
optical conductivity $\sigma _{\alpha \beta }(\omega )$ takes the
form\cite{Li2}
\begin{eqnarray}
\sigma _{\alpha \beta }(\omega ) &=&\frac{-i}{2\pi l_{B}^{2}}%
\sum_{N,N^{\prime },s,s^{\prime }}\frac{f_{N,s}-f_{N^{\prime },s^{\prime }}}{%
E_{N,s}-E_{N^{\prime },s^{\prime }}}  \notag \\
&&\times \frac{\langle N,s|j_{\alpha }|N^{\prime },s^{\prime }\rangle
\langle N^{\prime },s^{\prime }|j_{\beta }|N,s\rangle }{\omega
-E_{N,s}+E_{N^{\prime },s^{\prime }}+i/(2\tau )}  \label{CAB}
\end{eqnarray}%
where $1/\tau $ is a small constant residual scattering term and $f_{N,s}$
the Fermi-Dirac distribution function given by $1/(e^{\beta (\omega -\mu
)}+1)$ with $\beta $ the inverse temperature $T$ and $\mu $ the chemical
potential. Defining%
\begin{equation}
H(N,s,s^{\prime })=-1+s\sqrt{1/4+2NP}-s^{\prime }\sqrt{1/4+2(N+1)P}
\label{HN}
\end{equation}%
and%
\begin{eqnarray}
&&F(N,s,s^{\prime })=(\frac{\sqrt{N}}{\sqrt{2}}C_{\uparrow ,N+1,s^{\prime
}}^{\ast }C_{\uparrow ,N,s}+\frac{\sqrt{N+1}}{\sqrt{2}}  \notag \\
&&\times C_{\downarrow ,N+1,s^{\prime }}^{\ast }C_{\downarrow ,N,s}-\sqrt{P}%
C_{\uparrow ,N+1,s^{\prime }}^{\ast }C_{\downarrow ,N,s})^{2}  \label{FN}
\end{eqnarray}%
for $N\neq 0$ and for $N=0$
\begin{equation}
H(0,s)=-1/2-s\sqrt{1/4+2P}  \label{H0s}
\end{equation}%
\begin{equation}
F(0,s)=(\frac{1}{\sqrt{2}}C_{\downarrow ,1,s}-\sqrt{P}C_{\uparrow ,1,s})^{2}
\label{F0s}
\end{equation}

\begin{figure}[tp]
\begin{center}
\includegraphics[height=3.2in,width=3.5in]{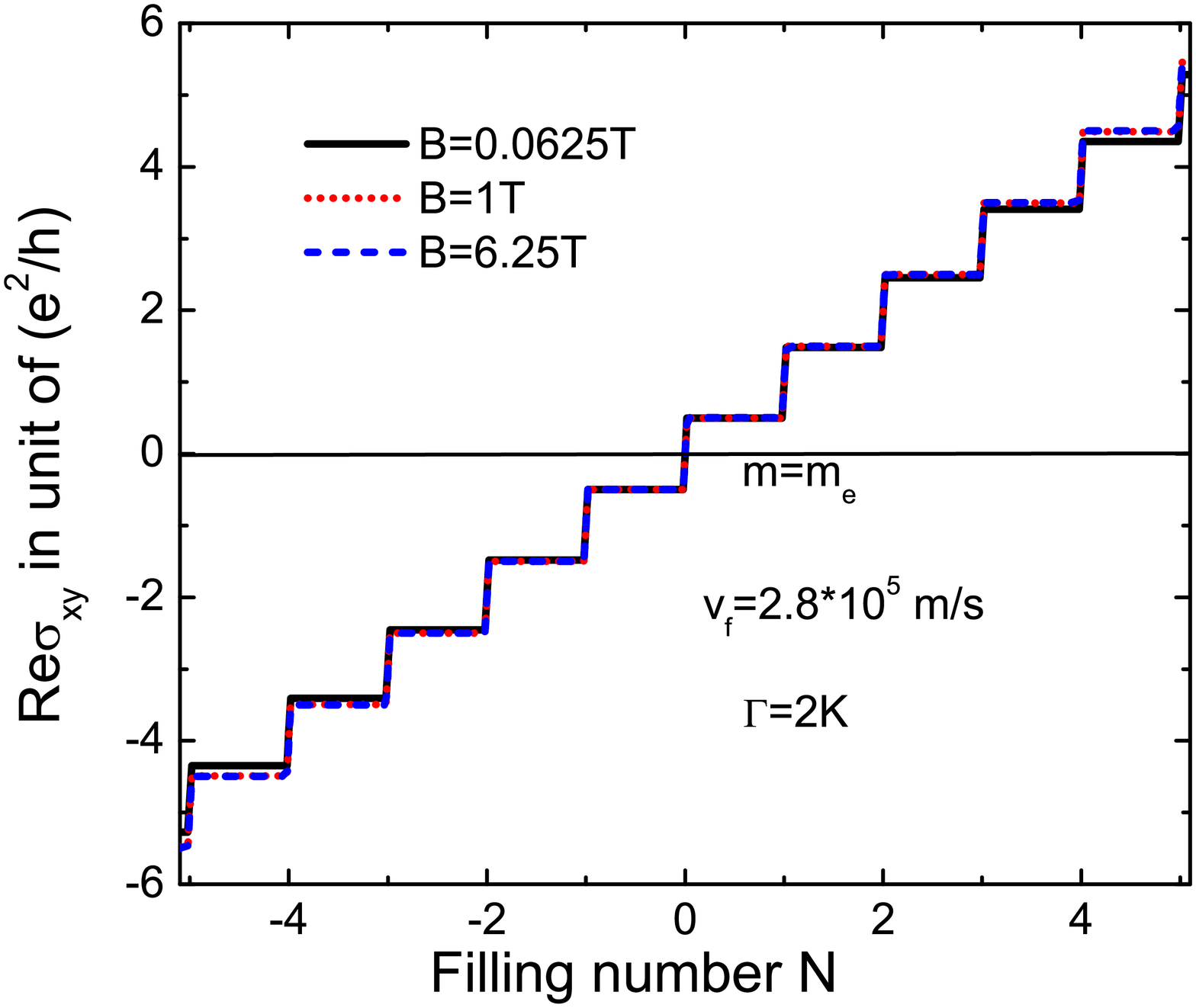}
\end{center}
\caption{(Color online) Real part of DC Hall conductivity $Re\protect\sigma %
_{xy}(\protect\omega =0)$ in units of $e^{2}/h$ as a function of
filling number N for three cases. The solid black steps apply to a
small magnetic field of $0.0625$ Tesla. It includes a small
Schr\"{o}dinger contribution to the Hamiltonian (1) with mass $m$
equal to the free electron mass in addition to a dominant Dirac
contribution with $v_{F}=2.8\ast 10^{5}$ m/s. It deviates only very
slightly from the pure Dirac case and the quantization is $1/2$,
$3/2$, $5/2$... on the vertical axis. For (red) dotted curve $B$ has
been increased to $1$ Tesla and for the (blue) dashed curve $B=6.25$
T. These curves do not deviate from Dirac quantization of the Hall
plateaux. } \label{fig3}
\end{figure}
the Kubo formula (\ref{CAB}) for the DC limit of the Hall
conductivity $\sigma _{xy}(\omega ) $ reduces to
\begin{eqnarray}
&&Re\sigma _{xy}=\frac{e^{2}}{h}\{\sum_{N=1,s,s^{\prime }}[\tanh \frac{%
(E_{N+1,s^{\prime }}-\mu )}{2T}  \notag \\
&&-\tanh \frac{(E_{N,s}-\mu )}{2T}]\frac{F(N,s,s^{\prime })}{%
H^{2}(N,s,s^{\prime })}+\sum_{s}\frac{F(0,s)}{H^{2}(0,s)}  \notag \\
&&\times \lbrack \tanh \frac{(E_{1,s}-\mu )}{2T}-\tanh \frac{(E_{0,+}-\mu )}{%
2T}]\}
\end{eqnarray}%
In these expressions $F(N,s,s^{\prime })$ and $H(N,s,s^{\prime })$ involve
the wave functions associated with the Hamiltonian (\ref{HM}) with $F$ and $%
H $ given by Eq.~(\ref{HN},\ref{FN},\ref{H0s},\ref{F0s}). Here we
refer to the combination $F/H^{2}$ as the optical matrix element
(OME).

\section{DC Hall conductivity in the Dirac limit and Schr\"{o}dinger limit}

In the pure Dirac limit there is no mass term, $P\rightarrow
\infty$, we have
\begin{eqnarray}
H(N,s,s^{\prime }) &=&s\sqrt{2NP}-s^{\prime }\sqrt{2(N+1)P}  \notag \\
F(N,s,s^{\prime }) &=&(-\sqrt{P}C_{\uparrow ,N+1,s^{\prime }}^{\ast
}C_{\downarrow ,N,s})^{2}
\end{eqnarray}%
\begin{equation}
C_{\uparrow ,N,s}\approx \frac{1}{\sqrt{2}},C_{\downarrow ,N,s}\approx \frac{%
s}{\sqrt{2}}
\end{equation}%
for $N\neq 0$ and for $N=0$
\begin{equation}
H(0,s)=-s\sqrt{2P},F(0,s)=(-\frac{\sqrt{P}}{\sqrt{2}})^{2}
\end{equation}%
Thus the DC Hall conductivity becomes%
\begin{eqnarray}
&&Re\sigma _{xy}(\omega )  \notag \\
&=&\frac{e^{2}}{4h}\{\sum_{N=1,s,s^{\prime }}\frac{(\tanh \frac{%
(E_{N+1,s^{\prime }}-\mu )}{2T}-\tanh \frac{(E_{N,s}-\mu )}{2T})}{(s\sqrt{2N}%
-s^{\prime }\sqrt{2(N+1)})^{2}}  \notag \\
&&+\sum_{s}(\tanh \frac{(E_{1,s}-\mu )}{2T}-\tanh \frac{(E_{0,+}-\mu )}{2T}%
)\}  \notag \\
&=&\frac{e^{2}}{4h}\{-2\tanh \frac{(E_{0,+}-\mu )}{2T}+\sum_{N=1}[\tanh
\frac{(E_{N,+}-\mu )}{2T}  \notag \\
&&+\tanh \frac{(E_{N,-}-\mu )}{2T}]L(N)\}
\end{eqnarray}%
where $L(N)=\frac{1}{(\sqrt{2N}-\sqrt{2(N-1)})^{2}}-\frac{1}{(\sqrt{2N}-%
\sqrt{2(N+1)})^{2}}+\frac{1}{(\sqrt{2N}+\sqrt{2(N-1)})^{2}}-\frac{1}{(\sqrt{%
2N}+\sqrt{2(N+1)})^{2}}=-2$. So we recover the well known result \cite%
{Gusynin2} which applies to pure relativistic Dirac fermions%
\begin{eqnarray}
&&Re\sigma _{xy}=\frac{e^{2}}{h}\{\frac{1}{2}\tanh \frac{\mu }{2T}+\frac{1}{2%
}\times  \notag \\
&&\sum_{N=1}[\tanh \frac{(\mu -E_{D,N})}{2T}+\tanh \frac{(\mu +E_{D,N})}{2T}%
]\}
\end{eqnarray}
where $E_{D,N}=\sqrt{2N\alpha ^{2}/l_{B}^{2}}$. This expression
appears as Eq.~(6) in the paper of Gusynin and Sharapov
\cite{Gusynin2} and gives the quantization series $1/2,3/2,5/2...$.
In the pure Schr\"{o}dinger limit, $P=0$, we have
\begin{eqnarray}
H(N,s,s^{\prime }) &=&-1  \notag \\
F(N,s,s^{\prime }) &=&(\frac{\sqrt{N+1}}{\sqrt{2}})^{2}=\frac{N+1}{2}
\end{eqnarray}%
so the DC Hall conductivity becomes
\begin{eqnarray}
Re\sigma _{xy} &=&\frac{e^{2}}{h}\sum_{N=0}[(\tanh \frac{(E_{S,N+1}-\mu )}{2T%
}  \notag \\
&&-\tanh \frac{(E_{S,N}-\mu )}{2T})(N+1)]  \label{Sch}
\end{eqnarray}%
where $E_{S,N}=\hbar ^{2}(N+\frac{1}{2})/(ml_{B}^{2})$. This is the
standard expression for the classical Schr\"{o}dinger case, and
gives the usual quantization $0,1,2,3...$ as discussed in
Ref.~[\onlinecite{Gusynin2}].

\section{Numerical Results}

\begin{figure}[tp]
\begin{center}
\includegraphics[height=3.0in,width=3.0in]{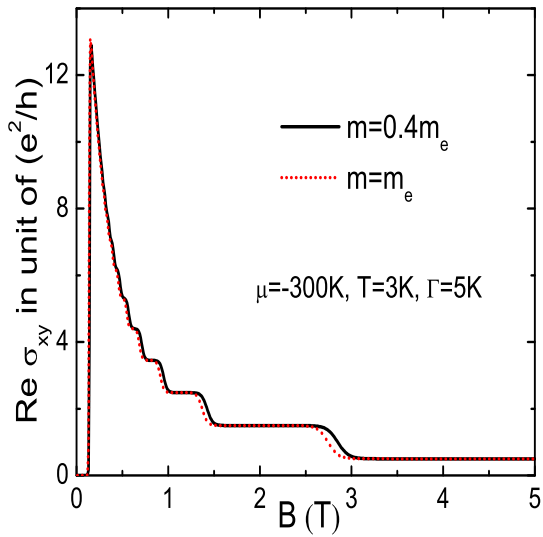}
\end{center}
\caption{(Color online) Real part of DC Hall conductivity $Re\protect\sigma %
_{xy}(\protect\omega=0)$ in units of $e^2/h$ as a function of magnetic field
$B$ in Tesla. The chemical potential is $\protect\mu=-300K$, the temperature
$T=3K$ and the residual scattering rate in Eq.~(\protect\ref{CAB}) is set at
$\Gamma=5K=1/(2\protect\tau)$. The sign of the Hall conductivity by our
definition is negative for $\protect\mu=-300K$, only in this figure do we
change the sign to be positive. The dotted (red) curve are results for the
mass $m=m_e$ (free electron mass) and the solid black for $m=0.4m_e$ which
increases the effect of the Schr\"{o}dinger term in the Hamiltonian (1). }
\label{fig4}
\end{figure}

\begin{figure}[tp]
\begin{center}
\includegraphics[height=3.2in,width=3.5in]{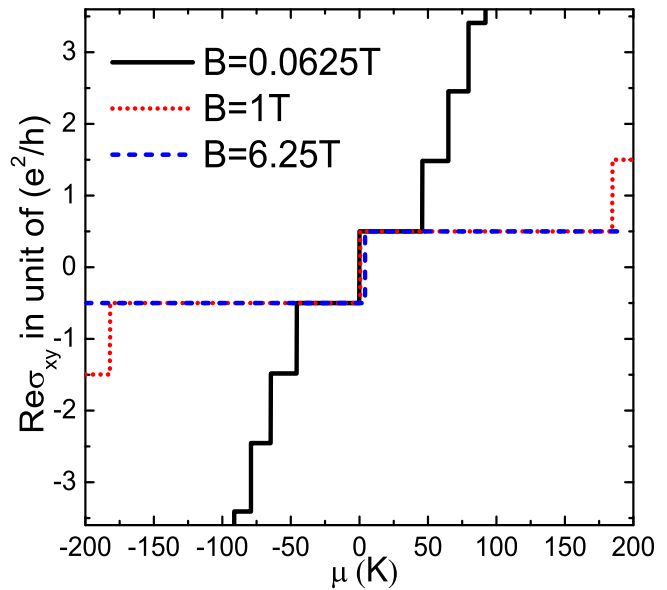}
\end{center}
\caption{(Color online) Real part of DC Hall conductivity $Re\protect\sigma %
_{xy}(\protect\omega=0)$ in units of $e^2/h$ as a function of chemical
potential ($\protect\mu$) in units of degree $K$. Here $v_F=2.8*10^5$ m/s
and $m=m_e$. Three values of magnetic field $B$ are considered, solid
(black) curve, $B=0.0625$ Tesla, dotted (red) curve, $B=1$ Tesla and short
dashed (blue) curve $B=6.25$ Tesla. In contrast to Fig.~\protect\ref{fig3}
where the horizontal axis is filling number and the steps occur at multiples
of one, there is now no quantization on $\protect\mu $ associated with the
various steps.}
\label{fig5}
\end{figure}

\begin{figure}[tp]
\begin{center}
\includegraphics[height=5in,width=3.5in]{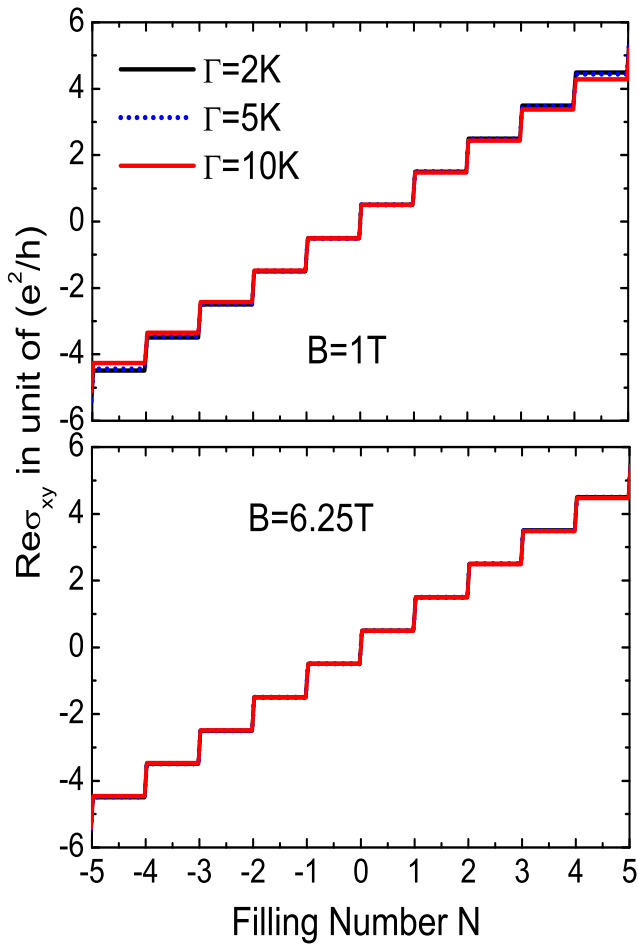}
\end{center}
\caption{(Color online) Real part of DC Hall conductivity $Re\protect\sigma %
_{xy}(\protect\omega =0)$ in units of $e^{2}/h$ as a function of filling
number N for magnetic field $B=1$ Tesla (top frame) and $B=6.25$ Tesla
(bottom frame). Here $v_F=2.8*10^5$ m/s and $m=m_e$. The solid black curve
is for the residual scattering rate in Eq.~(\protect\ref{CAB}) $\Gamma =1/(2%
\protect\tau)=2K$, dotted blue for $5K$ and solid red for $10K$. At
the smaller value of $B$ (upper frame) small differences in the
heights of the plateaux can be seen. These vanish as $B$ is
increased (lower frame). } \label{fig6}
\end{figure}

\begin{figure}[tp]
\begin{center}
\includegraphics[height=3in,width=3.5in]{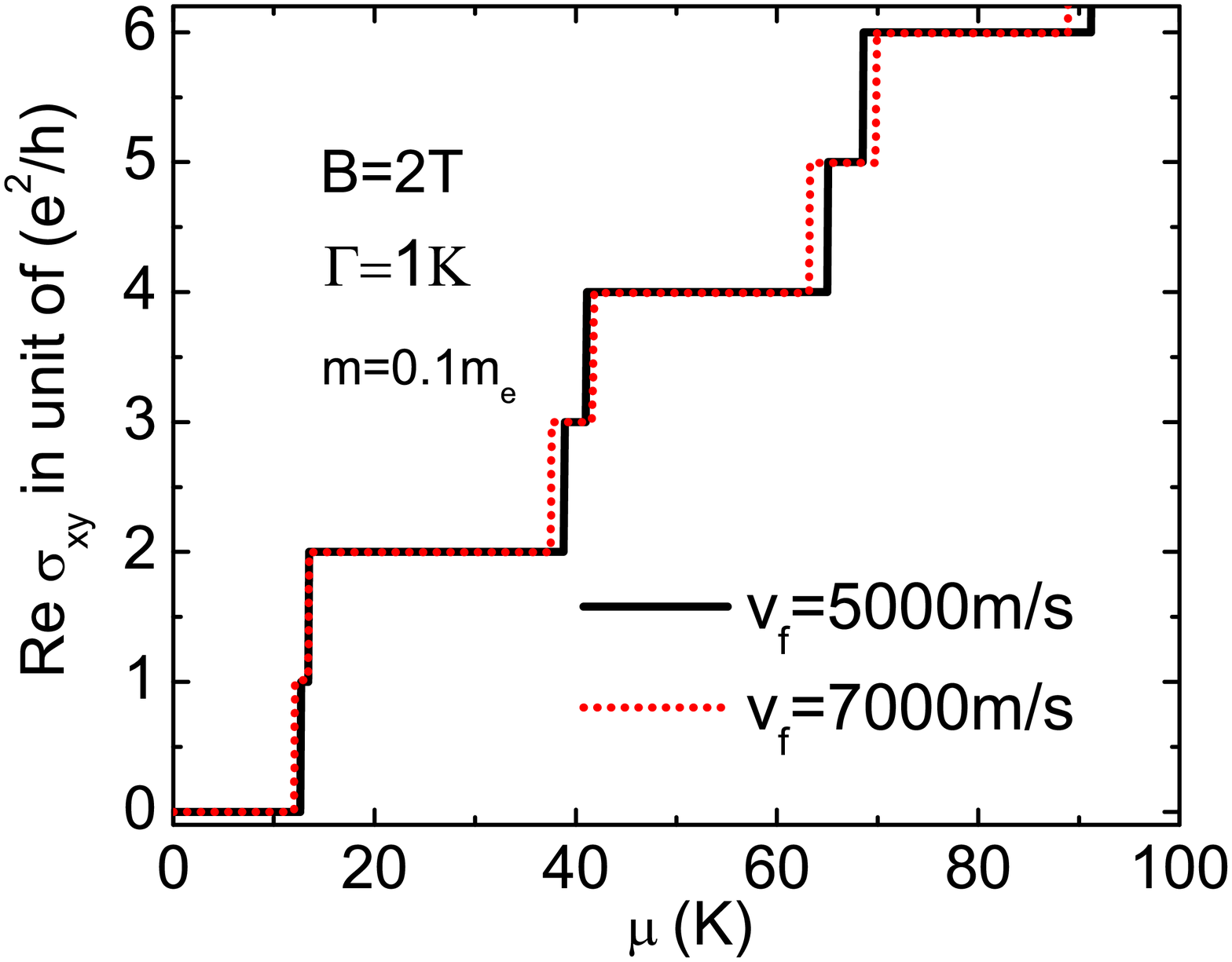}
\end{center}
\caption{(Color online) Real part of DC Hall conductivity $Re\protect\sigma %
_{xy}(\protect\omega =0)$ in units of $e^{2}/h$ as a function of chemical
potential $\protect\mu $ for magnetic field $B=2$ Tesla. The parameters used
\protect\cite{Fabian} are typical for spintronic semiconductors with Rashba
spin orbit coupling, and are the same as for Fig.~1b. }
\label{fig7}
\end{figure}
While we have just seen that our general expression for the integer quantum
Hall effect properly reduces to the well known results for relativistic and
classical electrons in the appropriate limits namely $P=\infty $ and $P=0$
respectively, what we are mainly interested in here, is the case of $1/P<<1$%
. We want to obtain a first correction to pure Dirac i.e. a first
correction in an expansion in powers of $1/P$. In the appendix we
show how this can be done. The first formula Eq.~(\ref{XY0}) applies
to any finite $P$ case and is the start of our analytic work which
ends with Eq.~(\ref{XYF}). This simple analytic formula gives
exactly the same results as does Eq.~(\ref{XY0}) provided $x=1/P$ is
small ($x<<1$). Note that the lowest power in $x$ to appear in
Eq.~(A2) to Eq.~(A7) which give the corrections to
the optical matrix elements, is $\sqrt{x}$ rather than $x$. By definition $%
\sqrt{x}=E_{0}/E_{1}=\frac{\hbar }{mv_{F}}\sqrt{e|B|/\hbar }$. This quantity
can be small for three reasons, $m$ can be made large which is equivalent to
having a very small Schr\"{o}dinger term in the Hamiltonian Eq.~(\ref{H0}).
The velocity $v_{F}$ large, also makes $\sqrt{x}$ small as it corresponds to
increasing the importance of the Dirac term. Finally for fix $m$ and $v_{F}$%
, $\sqrt{x}$ can still be made small by making the magnetic field
small. For the range of parameters used in this work the deviations
from pure Dirac behavior of the relativistic integer quantum Hall
effect is negligible except at small $B$ as we see in
Fig.~\ref{fig3}. The solid black curve
corresponds to a rather small value of magnetic field $B=0.0625T$ for which $%
\sqrt{x}=1/\sqrt{P}=0.0027$. The curve shows very small deviations from the
pure Dirac case as given for example in Refs.~[\onlinecite{Gusynin1},%
\onlinecite{Gusynin2}]. Here we remind the reader that the
parameters chosen as representative of a topological insulator were
$v_{F}=2.8\ast 10^{5}m/s$ and $m=m_{e}$ (the bare electron mass) and
the scattering rate $\Gamma =1/(2\tau )=2K$ in Eq.~(\ref{CAB}). It
is this $\Gamma $ which is responsible for the small changes of the
black curve from relativistic quantization which would certainly
apply in the clean limit where the Landau levels remain individually
well defined even for small values of the magnetic field $B$. If $B$
is increased to $B=1T$ (Tesla) which corresponds to a $\sqrt{x}$
value of $0.011$, the red dotted curve is exactly the pure Dirac
curve as is the blue dashed curve for $B=6.25T$.

While formula Eq.~(\ref{XY0}) is generally valid, it is hard to use
because it requires knowledge of $H(N,s,s^{\prime })$,
$F(N,s,s^{\prime })$ and $H(0,s)$, $F(0,s)$ of formulas
Eq.~(\ref{HN}) to Eq.~(\ref{F0s}) respectively. By contrast the
final but approximate formula obtained in appendix A
\begin{eqnarray}
&&Re\sigma _{xy}=\frac{e^{2}}{h}\{\frac{1}{2}\tanh \frac{(\mu -E_{0,+})}{2T}+%
\frac{1}{2}\times   \notag \\
&&\sum_{N=1}[\tanh \frac{(\mu -E_{N,+})}{2T}+\tanh \frac{(\mu -E_{N,-})}{2T}%
]\}  \label{final}
\end{eqnarray}%
can be evaluated directly from a knowledge of $\mu $ (the chemical
potential) and the energies of the LL $E_{N,s}$. There are no
corrections in this equation coming from the optical matrix elements
(A2) to (A7). These were expanded to order $x^{3/2}$ and found to
cancel out entirely. What could provide deviations of the Hall
plateaux from the pure relativistic Dirac prediction are terms in
Eq.~(\ref{final}) proportional to $\sqrt{x}$ and $x$, and higher
orders that appear in the energies. But these appear in the $\tanh
$'s which at zero temperature are just step functions one or zero
and consequently this does not change the height of the Hall
plateaux as we next further emphasize. However it does change the
value of the chemical potential at which the jumps occur.

In Fig.~\ref{fig4} we show our results for the DC Hall conductivity $%
Re\sigma _{xy}(\omega =0)$ in units of $e^{2}/h$ as a function of magnetic
field $B$ in Tesla for fixed value of chemical potential. We have chosen the
other parameters to correspond to those used in Fig.3a of Ref.~[%
\onlinecite{Gusynin1}] where the case of graphene was discussed and this
serves as a test of the accuracy of our work. The chemical potential is $\mu
=-300K$, the temperature $T=3K$, the residual scattering rate $\Gamma =5K$.
The dotted curve (red) includes an effective mass ($m$) equal to the free
electron mass ($m_{e}$) and the solid (black) curve has $m=0.4m_{e}$ which
implies a relatively larger Schr\"{o}dinger contribution and consequently
larger deviations from a pure Dirac case. Accounting for a degeneracy factor
of $4$ (spin and valley) used in Ref.~[\onlinecite{Gusynin1}] and not in our
work, our results agree with those presented in their Fig.3a when we
consider pure graphene. As we expect, there are however small deviations
between our results and those of Ref.~[\onlinecite{Gusynin1}] which become
more noticeable as $m$ decreases. It is important to compare carefully the
results of Fig.~\ref{fig4} with those of Fig.~\ref{fig3}. In Fig.~\ref{fig4}
we have kept the chemical potential fixed at a value of $\mu =-300K$ and
varied $B$. The step seen in the dotted curve around $B\approx 3$ Tesla
corresponds to the case when the $N=0$ LL is about to cross through the
chemical potential and so this corresponds to the first step in Fig.~\ref%
{fig3} rather than to a large filling N value. We emphasize that
there is no change in the quantization of $\sigma _{xy}(\omega =0)$
from the pure Dirac case but there is a change in the value of $B$
where the steps occur and this is largest at large $B$. As an
example we see a shift of $4\%$ in the first step which is reduced
to $2.6\%$ in the second step (smaller $B$ value). Finally the
impurity scattering embodied in $\Gamma $ and finite temperature has
smeared out the steps between plateaux as we expect. Also, the range
of $B$ over which a given plateau manifests is greatly reduced as
$B$ is reduced. This is to be contrasted with the results in
Fig.~\ref{fig3} where the distance between plateaux along the
horizontal axis is uniform.

If instead of the filling number we had used the chemical potential
for the horizontal axis as we show in Fig.~\ref{fig5}, we no longer
get equal distance steps. For the smallest value of $B=0.0625$ T
solid (black) curve the steps from one plateau to the next are
narrow and become even narrower as $\mu $ increases because the
spacing in energy between LL decreases with increasing energy. But
this spacing also increases with increasing $B$ as can be seen in
the dotted (red) curve for $B=1$ Tesla and the dashed (blue) curve
for $B=6.25$ Tesla. For this last curve the second step falls
outside the range of $\mu $ shown in the diagram. Another point to
be made, which is seen most clearly in this last curve, is that the
first step when the Hall conductivity goes from $-1/2$ to $1/2$ no
longer occurs exactly at zero value of chemical potential but is
rather at $\mu=E_{N=0}=\hbar^2/2ml_{B}^{2}$. This is a
characteristic difference between pure Dirac and a TI. This
important feature is lost in Fig.3 where filling number is used
instead of the chemical potential on the horizontal axis. Finally in
Fig.~\ref{fig6} we show results for three values of residual
scattering namely $\Gamma =2K$ solid (black), $\Gamma =5K$ dashed
(blue) and $\Gamma =10K$ solid (red). The top frame is for $B=1$
Tesla the bottom for $B=6.25$ Tesla. In all cases we see only a
small effect of $\Gamma $ on the quantization which increases
slightly with increasing filling number and decreases with
increasing value of $B$.

It is also of interest to consider the opposite limit when the dominant
magnetic energy comes from the Schr\"{o}dinger term in (1) and the Dirac
contribution provides a small correction. In that limit the appropriate
expansion parameters is $P=(E_{1}/E_{0})^{2}$, and the general formula (A1)
is to be expanded in powers of $P$ with $P<<1$. For $P=0$ we have already
seen that it reduces to the classical non relativistic case. Keeping a first
correction to account for a small Dirac contribution in (1) we obtain
\begin{eqnarray}
&&Re\sigma _{xy}  \notag \\
&=&\frac{e^{2}}{h}\{\sum_{N=0}[\tanh \frac{(E_{N+1,+}-\mu )}{2T}-\tanh \frac{%
(E_{N,+}-\mu )}{2T}]  \notag \\
&&\times \lbrack \frac{1+N}{2}-(1+N)P+6(1+N)^{2}P^{2}]  \notag \\
&&+\sum_{N=1}[\tanh \frac{(E_{N+1,-}-\mu )}{2T}-\tanh \frac{(E_{N,-}-\mu )}{%
2T}]  \notag \\
&&\times \lbrack \frac{N}{2}+NP-6N^{2}P^{2}]  \notag \\
&&+\sum_{N=0}[\tanh \frac{(E_{N+1,-}-\mu )}{2T}-\tanh \frac{(E_{N,+}-\mu )}{%
2T}]  \notag \\
&&\times \lbrack P-6(1+2N)P^{2}]\}
\end{eqnarray}

This equation contains powers of $P$ and $P^{2}$ terms from the
expansion of the optical matrix elements. However as we will see
below these terms drop out of the final formula (27). This means
that corrections to the classical case coming from the OME are very
small and must be of higher order than $(E_{1}/E_{0})^{4}$. This is
to be contrasted with the case applicable for topological insulators
where we found that the correction to the pure relativistic case
must be of higher order than cubic in $E_{1}/E_{0}$.

To understand better the meaning of Eq. (25) in the limit $P<<1$ we
begin by expanding the LL energies of Eq. (4) in powers of $P$, we
get for $N\neq 0$,
\begin{equation}
E_{N,s}=E_{0}(N+s/2)+s2N(E_{1}/E_{0})^{2}E_{0}
\end{equation}%
and $E_{N=0}$ retains the form it has in Eq. (5). If first we neglect the $%
(E_{1}/E_{0})^{2}$ correction in Eq. (26) we see that for positive $s$, $%
E_{0}$ plus the sequence $N=1,2...$ give the classical result for the LL
series and for negative $s$, $N=1,2...$ give a second such sequence. Thus we
have two LL sequences which accounts for spin degeneracy. When $%
(E_{1}/E_{0})^{2}<<1$ the negative $s$ sequence is slightly shifted down and
the positive $s$ sequence is shifted up by the same amount and $%
E_{N+1,-}=E_{N,+}-2(2NE_{0}P)$. The two sequence involved can be reorganized
to get
\begin{eqnarray}
&&Re\sigma _{xy}  \notag \\
&=&\frac{e^{2}}{h}\{\sum_{N=0}[\tanh \frac{(E_{N+1,+}-\mu )}{2T}-\tanh \frac{%
(E_{N,+}-\mu )}{2T}]  \notag \\
&&\times (1+N)/2+\sum_{N=1}[\tanh \frac{(E_{N+1,-}-\mu )}{2T}  \notag \\
&&-\tanh \frac{(E_{N,-}-\mu )}{2T}]\times \frac{N}{2}\}
\end{eqnarray}%
This is a second important result of this work.

In Fig.~\ref{fig7} we show
results for the DC Hall conductivity as a function of chemical potential $%
\mu $ for parameters typical of present day spintronic semiconductors. \cite%
{Fabian} The Schr\"{o}dinger mass $m$ in Eq. (1) is set at ($m=0.1m_{e}$)
one tenth of the bare electron mass and two values of Dirac velocities
namely $v_{F}=5000m/s$ (solid black curve) and $v_{F}=7000m/s$ (dotted red)
are considered. The residual scattering rate $\Gamma =1K$ and the magnetic
field is $B=2T$ for both cases. Even though we have taken values of the
Dirac velocity which are near their maximum in spintronic semiconductors,%
\cite{Fabian} we see that the deviations from the classical case are small.
The quantization remains classical but two such series are involved which
are slightly shift with the shift between the two increasing with increasing
$\mu $ as we expect from Eq. (26).

\section{Summary and Conclusions}

The helical electrons which exist at the surface of topological insulators
have electronic dispersion curves which include a subdominant Schr\"{o}%
dinger quadratic in momentum part characterized by an effective mass $m$ and
a dominant Dirac linear in momentum part described by a fermi velocity $v_{F}
$. The small quadratic piece distorts the usual Dirac cones of graphene and
gives them instead an hourglass shape and is responsible for particle-hole
asymmetry. In a magnetic field oriented perpendicular to the plane of the
helical surface electrons, Landau levels form but these require a much more
complicated mathematical description than when either Schr\"{o}dinger or
Dirac term is present separately. We have derived formulas for the DC Hall
conductivity that cover the mixed case and which are valid for any value of
Schr\"{o}dinger and Dirac energy scale, $E_{0}=\hbar e|B|/m$ and $%
E_{1}=\alpha \sqrt{e|B|/\hbar }$ respectively. In general our
formulas need to be evaluated numerically, as the matrix elements of
the current now do not have a simple form. The Landau level energies
are also complicated expressions of the LL index N. In the limit
$E_{0}=0$, $\sigma _{xy}(\omega =0)$ reduces to the
known quantized Hall plateaux $1/2$, $3/2$, $5/2...$ (Dirac) in units of $%
e^{2}/h$ while for $E_{1}=0$ the plateaux are at $0$, $1$, $2$, $3...$ (Schr\"{o}%
dinger).

We have also reduced our general expressions to a much simpler form
in the limit when $E_{0}$ can be considered to be a small
perturbation on the pure Dirac case. This is the case of greatest
interest in this paper. To accomplish this we expanded the optical
matrix elements to second order of perturbation theory
in powers of $\sqrt{x}=(E_{0}/E_{1})$. This leads to a simple formula for $%
Re\sigma _{xy}(\omega =0)$ which depends only on the Landau level energies $%
E_{N,s}$ with $s=\pm 1$, given by Eq.~(\ref{landau}) and Eq.~(\ref{landau0}%
), and on the chemical potential ($\mu $). The parameter $x$ coming
from the expansion of the optical matrix element dropped out
entirely. This formula is given by Eq.~(\ref{final}) which is one of
our important results. It has the same form as for the pure Dirac
case except that it is the Landau level energies of the TI which
appear in the thermal factors and these contain a contribution from
the subdominant Schr\"{o}dinger term in the Hamiltonian (1). The
Hall plateaux however keep their relativistic quantization even
though $x$ is not zero, and the value of chemical potential at which
the hall conductivity jumps from a negative to a positive value is
no longer zero but is at $\mu=E_0/2$.

The central parameter $\sqrt{x}=E_{0}/E_{1}=\sqrt{e|B|\hbar }/(mv_{F})$ can
be small for three reasons. The magnetic field can be made small, or the Schr%
\"{o}dinger mass $m$ or the Dirac Fermi velocity ($v_{F}$) can be
made large. A large mass means a small quadratic term in our
Hamiltonian (1) and a large $v_{F}$ means a large Dirac
contribution. Our numerical work based on the exact equations for
the DC Hall plateaux confirms that for a large range of $x$, we
indeed recover the pure Dirac quantization pattern as our Eq. (24)
predicts. These results are of interest within the context of
presently discovered topological insulators. We have also considered the opposite limit when the Schr\"{o}%
dinger term dominates and the Dirac term is a small correction. In this case
the appropriate expansion of the optical matrix elements which appear in our
general formula is to consider powers of $E_{1}/E_{0}$. Working to order $%
(E_{1}/E_{0})^{4}$ we find a complete cancelation of these factors
in the optical matrix elements and we are left with the classical
quantization series for the Hall plateaux split however into two
series with splitting related to shifts in the Landau levels
energies brought about by the subdominant spin orbit coupling. This
is another of our important results and is relevant to present day
spintronic semiconductors.

\begin{acknowledgments}
This work was supported by the Natural Sciences and Engineering
Research Council of Canada (NSERC) and the Canadian Institute for
Advanced Research (CIFAR).
\end{acknowledgments}

\appendix

\section{Derivation of the DC hall conductivity for the topological insulator%
}

With the mass term included the Hall conductivity is given by
\begin{eqnarray}
&&Re\sigma _{xy}=\frac{e^{2}}{h}\{\sum_{N=1}[(\tanh \frac{(E_{N+1,+}-\mu )}{%
2T}  \notag \\
&&-\tanh \frac{(E_{N,+}-\mu )}{2T})\frac{F(N,+,+)}{H^{2}(N,+,+)}  \notag \\
&&+(\tanh \frac{(E_{N+1,+}-\mu )}{2T}-\tanh \frac{(E_{N,-}-\mu )}{2T})\frac{%
F(N,-,+)}{H^{2}(N,-,+)}  \notag \\
&&+(\tanh \frac{(E_{N+1,-}-\mu )}{2T}-\tanh \frac{(E_{N,+}-\mu )}{2T})\frac{%
F(N,+,-)}{H^{2}(N,+,-)}  \notag \\
&&+(\tanh \frac{(E_{N+1,-}-\mu )}{2T}-\tanh \frac{(E_{N,-}-\mu )}{2T})\frac{%
F(N,-,-)}{H^{2}(N,-,-)}]  \notag \\
&&+\sum_{s}(\tanh \frac{(E_{1,s}-\mu )}{2T}-\tanh \frac{(E_{0,+}-\mu )}{2T})%
\frac{F(0,s)}{H^{2}(0,s)}\}  \label{XY0}
\end{eqnarray}%
In a general case Eq.~(A1) is complicated, although it is explicit,
because the Landau level energies are not simple functions of the LL
index N and more importantly the current matrix elements are
particularly long algebraic expressions. After expanding the OME in
power of $x=1/P$ and retaining terms to the order of $x^{3/2}$ only,
we obtained
\begin{eqnarray}
&&\frac{F(N,+,+)}{H^{2}(N,+,+)}=\frac{1}{8(\sqrt{N}-\sqrt{1+N})^{2}}  \notag
\\
&&+\frac{(1/\sqrt{N}+1/\sqrt{N+1})\sqrt{x}}{16\sqrt{2}}  \notag \\
&&-\frac{(1/\sqrt{N^{3}}+1/\sqrt{(N+1)^{3}})x^{3/2}}{256\sqrt{2}}\text{,}
\end{eqnarray}%
\begin{eqnarray}
&&\frac{F(N,-,+)}{H^{2}(N,-,+)}=\frac{1}{8(\sqrt{N}+\sqrt{1+N})^{2}}  \notag
\\
&&-\frac{(1/\sqrt{N}-1/\sqrt{N+1})\sqrt{x}}{16\sqrt{2}}  \notag \\
&&+\frac{(1/\sqrt{N^{3}}-1/\sqrt{(N+1)^{3}})x^{3/2}}{256\sqrt{2}}\text{,}
\end{eqnarray}%
\begin{eqnarray}
&&\frac{F(N,+,-)}{H^{2}(N,+,-)}=\frac{1}{8(\sqrt{N}+\sqrt{1+N})^{2}}  \notag
\\
&&+\frac{(1/\sqrt{N}-1/\sqrt{N+1})\sqrt{x}}{16\sqrt{2}}  \notag \\
&&-\frac{(1/\sqrt{N^{3}}-1/\sqrt{(N+1)^{3}})x^{3/2}}{256\sqrt{2}}
\end{eqnarray}%
\begin{eqnarray}
&&\frac{F(N,-,-)}{H^{2}(N,-,-)}=\frac{1}{8(\sqrt{N}-\sqrt{1+N})^{2}}  \notag
\\
&&-\frac{(1/\sqrt{N}+1/\sqrt{N+1})\sqrt{x}}{16\sqrt{2}}  \notag \\
&&+\frac{(1/\sqrt{N^{3}}+1/\sqrt{(N+1)^{3}})x^{3/2}}{256\sqrt{2}}
\end{eqnarray}%
and
\begin{equation}
\frac{F(0,+)}{H^{2}(0,+)}=\frac{1}{4}+\frac{\sqrt{x}}{8\sqrt{2}}-\frac{%
x^{3/2}}{128\sqrt{2}}
\end{equation}%
\begin{equation}
\frac{F(0,-)}{H^{2}(0,-)}=\frac{1}{4}-\frac{\sqrt{x}}{8\sqrt{2}}+\frac{%
x^{3/2}}{128\sqrt{2}}
\end{equation}%
Substituting this into Eq.~(\ref{XY0}) leads to the approximate
expression for the Hall conductivity
\begin{eqnarray}
&&Re\sigma _{xy}=\frac{e^{2}}{h}\{\frac{1}{2}\tanh \frac{(\mu -E_{0,+})}{2T}+%
\frac{1}{2}\times   \notag \\
&&\sum_{N=1}[\tanh \frac{(\mu -E_{N,+})}{2T}+\tanh \frac{(\mu -E_{N,-})}{2T}%
]\}  \label{XYF}
\end{eqnarray}%
which is a central results of this work. It provides a simple compact
analytic formula for the DC Hall conductivity of a topological insulator
which includes a first correction to a dominant Dirac Hamiltonian with an
additional small subdominant Schr\"{o}dinger part i.e. a small piece
quadratic in momentum. In mathematical term we have made an expansion of the
optical matrix elements of Eq.~(\ref{XY0}), which is itself valid for any
value of Schr\"{o}dinger and Dirac, in power of $x=1/P$ where $P$ is the
Diracness defined as $E_{1}^{2}/E_{0}^{2}$. Here $E_{0}=\hbar e|B|/m$ and $%
E_{1}=\alpha \sqrt{e|B|/\hbar }$. This first energy $E_{0}$
determines the Landau levels for a pure Schr\"{o}dinger (classical)
case while $E_{1}$ is the magnetic energy associated with the LL for
a pure Dirac (relativistic) spectrum. All optical matrix element
corrections have dropped out to order $x^{3/2}$. Equation (A8)
differs from (21) for pure Dirac only through the appearance of the
energies $E_{0,+}$, $E_{N,+}$ and $E_{N,-}$ which here
include a small Schr\"{o}dinger piece. However at zero temperature the $%
\tanh $ factors in (A8) are just step functions which are either
zero or one as before but with the chemical potential value at which
these jumps occur modified by the Schr\"{o}dinger contribution to
the energies.

\end{document}